\documentclass[titlepage]{article}\usepackage[]{graphicx}\usepackage[]{color}
\makeatletter
\def\maxwidth{ %
  \ifdim\Gin@nat@width>\linewidth
    \linewidth
  \else
    \Gin@nat@width
  \fi
}
\makeatother

\definecolor{fgcolor}{rgb}{0.345, 0.345, 0.345}

\usepackage{framed}
\makeatletter
 {\par\unskip\endMakeFramed%
 \at@end@of@kframe}
\makeatother

\definecolor{shadecolor}{rgb}{.97, .97, .97}
\definecolor{messagecolor}{rgb}{0, 0, 0}
\definecolor{warningcolor}{rgb}{1, 0, 1}
\definecolor{errorcolor}{rgb}{1, 0, 0}

\usepackage{alltt}
\usepackage{graphicx}
\usepackage{paralist}
\usepackage{amsmath}
\usepackage{amssymb}
\usepackage{amsfonts}
\usepackage{adjustbox}
\usepackage{setspace}
\usepackage{color, colortbl}
\usepackage{mathtools}
\usepackage[dvipsnames]{xcolor}
\usepackage{array}
\usepackage{comment}

\definecolor{ClusterOne}{rgb}{2, 212, 9}

\usepackage{hyperref} 
\hypersetup{colorlinks=true,%
            linkcolor=black,%
            citecolor=black,%
            filecolor=black,%
            urlcolor=black,%
            linktoc=all
}

\usepackage{rotating} 
\usepackage[ruled,vlined]{algorithm2e} 

\usepackage{natbib}
\setcitestyle{authoryear,open={(},close={)}} 

\usepackage{parskip}
\DeclareGraphicsExtensions{.pdf,.png,.jpg}

\IfFileExists{upquote.sty}{\usepackage{upquote}}{}
\begin{document}
\title{Know Your Clients' behaviours: a cluster analysis of financial transactions}
\author{John R.J. Thompson, Longlong Feng, R. Mark Reesor, Chuck Grace} 
\maketitle

\newpage
John R.J. Thompson \emph{(Corresponding Author)}

Department of Statistical and Actuarial Sciences

The University of Western Ontario

London, Ontario N6A 5B7

jthomp83@uwo.ca
\\[3\baselineskip]
Longlong Feng

Department of Mathematics

Wilfrid Laurier University

Waterloo, Ontario N2L 3C5

feng0290@mylaurier.ca
\\[3\baselineskip]
R. Mark Reesor

Department of Mathematics

Wilfrid Laurier University

Waterloo, Ontario N2L 3C5

mreesor@wlu.ca
\\[3\baselineskip]
Chuck Grace

Department of Finance

Ivey Business School

London, Ontario N6G 0N1

cgrace@ivey.ca

\begin{abstract}
In Canada, financial advisors and dealers are required by provincial securities commissions and self-regulatory organizations--charged with direct regulation over investment dealers and mutual fund dealers--to respectively collect and maintain Know Your Client (KYC) information, such as their age or risk tolerance, for investor accounts. With this information, investors, under their advisor's guidance, make decisions on their investments which are presumed to be beneficial to their investment goals. Our unique dataset is provided by a financial investment dealer with over 50,000 accounts for over 23,000 clients. We use a modified behavioural finance recency, frequency, monetary model for engineering features that quantify investor behaviours, and machine learning clustering algorithms to find groups of investors that behave similarly. We show that the KYC information collected does not explain client behaviours, whereas trade and transaction frequency and volume are most informative. We believe the results shown herein encourage financial regulators and advisors to use more advanced metrics to better understand and predict investor behaviours.
  \\[3\baselineskip]
\end{abstract}


\doublespacing

\section{Introduction} \label{sec:introduction}
Investors hire financial advisors to help them select, facilitate, and manage their investment choices. In Canada, the client-advisor relationship varies by institution and regulatory regime. Some investors ask advisors to provide advice but ultimately make their own investment choices, other investors ask for a recommendation and then approve the advisor’s investment choices, while still others delegate full discretionary investment choices to the advisor.  However, regardless of the relationship, advisors are expected to provide recommendations that are suitable for the client.   

Suitability is described by regulators in Canada as a “meaningful dialogue with the client to obtain a solid understanding of the client's investment needs and objectives, and to explain how a proposed investment strategy is suitable for the client in light of the client's investment needs and objectives” \citep{ont14}. One of the suitability determinants for advisors is to determine the general investment needs and objectives of their client and any other factors necessary for them to determine whether a proposed purchase or sale is suitable (Know Your Client or KYC). The assumption is that any subsequent purchases or sales (trading behaviour) will conform to the KYC attributes and therefore be suitable\footnote{An important aspect of suitability is the product recommendation or KYP which we will address in subsequent papers.}.

In this paper, we consider unique interconnected datasets of financial transactions and KYC attributes to examine the relationship between KYC and trading behaviour. The KYC data is comprised of objective demographic and identifying information and subjective financial situation information, where both are used to generate a client's risk tolerance. We quantify trading behaviour through metrics designed using an extended Recency, Frequency, and Monetary (RFM) model from behavioural finance. Our hypothesis is that groups of investors with similar KYC attributes will have the same risk tolerance and trading behaviours. KYC information should inform a risk tolerance score which the financial advisor -- informed by suitability regulations -- uses to delineate client investment transactions.

We conduct our analysis using a machine learning $k$-prototypes clustering algorithm and visualize the clusters using $t$-distributed stochastic neighbour embeddings. Using advanced data analytics, our analysis shows that:

\begin{itemize}
\item Objective and subjective KYC data have little influence on trading behaviours (cf. Table \ref{tbl:overallfindings}). 
\item The distribution of risk tolerance across each clusters' trading behaviour is found to be similar, showing that trading behaviours may on occasion be inconsistent with the KYC generated risk tolerance (cf. Table \ref{tbl:overallfindings} and Figure \ref{fig:riskToleranceScoresByCluster}). 
\item KYC criteria appear to concentrate investors within narrow and rigid ‘swim lanes’ and appear to do a poor job of accommodating trading behaviours to the extremes--either highly risk-averse investors or those seeking higher risks (cf. Table \ref{tbl:overallfindings} and Figure \ref{fig:riskToleranceScoresByCluster}).
\end{itemize}

At the onset, the hypothesis for this paper was that a thorough and complete assessment of investor KYC data should lead to an accurate determination of risk tolerance and suitability requirements. In turn, those determinations should manifest downstream in trading behaviour and, eventually, in portfolio construction\footnote{In this paper we have focused on trading behaviour but we plan to address portfolio construction, asset mix, and risk and returns in subsequent papers.} and investment outcomes. 

\begin{figure}[!htbp]
\begin{centering}
  \includegraphics[width=18cm, keepaspectratio]{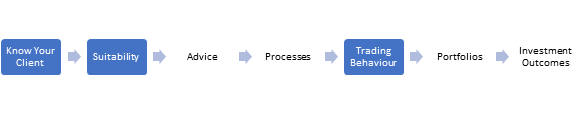}
  \caption{The downstream footprints of KYC regulations.}
  \label{fig:flowchart}
  \end{centering}
\end{figure}
 
Our conclusion that KYC data does not demonstrate a strong relationship to the trading behaviours exhibited by investors is important because ``Know Your Client" is a foundational principle behind the concept of ``suitability" and the corresponding investment regulatory framework deployed in many jurisdictions\footnote{See Proposed Amendments to National Instrument 31-103 Registration Requirements, Exemptions and Ongoing Registrant Obligations, December 2019 for a full discussion of the topic in Canada.}. The principle has also become more important as employers and governments de-risk retirement and savings programs post-2009 and move more of the burden of investment decision making from professional portfolio managers to individual investors\footnote{Pension coverage in Canada, January 2018, www150.statcan.gc.ca/n1/pub/75-006-x/2014001/article/14120-eng.htm.}. Furthermore, the topic has become more urgent given the events of early 2020. 

\begin{table}[]
\footnotesize
\caption{KYC demographics and trading behaviours compared to expected risk tolerance and anticipated risk tolerance for each cluster.}
\label{tbl:overallfindings}
\hspace*{-1.8cm} 
\begin{tabular}{p{2cm}p{3cm}p{3cm}p{3cm}p{3cm}p{3.5cm}}
 &  &  & \multicolumn{1}{c}{Clusters} &  &  \\
\multicolumn{1}{p{2cm}|}{Client trait} & \multicolumn{1}{p{3cm}|}{1 -- Active Traders} & \multicolumn{1}{p{3cm}|}{2 -- Early Savers} & \multicolumn{1}{p{3cm}|}{3 -- Just-in-Time} & \multicolumn{1}{p{3cm}|}{4 -- Older Investors} & 5 -- Systematic Savers \\ \hline
\multicolumn{1}{p{1.5cm}|}{KYC} & \multicolumn{1}{p{3cm}|}{Average age, income \& demographics. Average investment knowledge. Average \$ accounts \& balances} & \multicolumn{1}{p{3cm}|}{Slightly younger but average income \& demographics. Average investment knowledge. Average \$ accounts \& balances} & \multicolumn{1}{p{3cm}|}{Average age, income \& demographics. Average investment knowledge. Average \$ accounts \& balances} & \multicolumn{1}{p{3cm}|}{Older but average, income \& demographics. Average investment knowledge. Average \$ accounts \& balances} & Average age, income \& demographics. Average investment knowledge. Average \$ accounts \& balances \\ \hline
\multicolumn{1}{p{1.5cm}|}{Trade behaviour} & \multicolumn{1}{p{3cm}|}{Trade frequently in large amounts and appear sensitive to market influences} & \multicolumn{1}{p{3cm}|}{Smaller, regular deposits making use of PACs} & \multicolumn{1}{p{3cm}|}{Infrequent trades at seemingly random intervals} & \multicolumn{1}{p{3cm}|}{Primarily withdrawals, dividends, and interest payments} & Larger, systematic trades and re-balancing \\ \hline
\multicolumn{1}{p{1.5cm}|}{Risk tolerance observed average\footnotemark} & \multicolumn{1}{p{3cm}|}{3.19/5} & \multicolumn{1}{p{3cm}|}{3.18/5} & \multicolumn{1}{p{3cm}|}{3.12/5} & \multicolumn{1}{p{3cm}|}{2.95/5} & 3.19/5 \\ \hline
\multicolumn{1}{p{1.5cm}|}{Risk tolerance anticipated} & \multicolumn{1}{p{3cm}|}{5/5} & \multicolumn{1}{p{3cm}|}{4/5} & \multicolumn{1}{p{3cm}|}{3/5} & \multicolumn{1}{p{3cm}|}{1/5} & 2/5
\end{tabular}
\end{table}
\footnotetext{On a scale of 1 to 5 where 1 is a low or preservation risk tolerance and 5 is high or aggressive.}

At this point, it is important to acknowledge that investor behaviour is a complex and dynamic topic. Investor behaviour is not only driven by the investor’s personal motives such as their goals and financial needs but it is also influenced by the advisor relationship, dealer processes, regulatory obligations, and market influences. As well, while the client onboarding and discovery process is foundational, it is also contextual and time-dependent since the corresponding product recommendations are constantly changing in real-time. While the dataset and analysis used in this paper are unique, we are not privy to some of the subjective or undocumented influences and we cannot include them in our algorithms. We have also examined only one set period of time. It is therefore impossible for us to determine why the KYC process is not leading to the outcomes we would expect. Our analysis has inspired the question ``Could protocols be improved?” but we can’t answer the question without further research\footnotemark.
\footnotetext{Please refer to Section \ref{sec:discussion} for our future research plans.}

The paper reads as follows: The rest of Section \ref{sec:introduction} is a literature review on KYC regulations and trading behaviour and Section \ref{sec:featureEngine} introduces the client and advisor financial data collected by a dealer, and develops the features that were used to measure client behaviours. Section \ref{sec:clustering} describes the machine learning methods used to identify investor groups based on their KYC information and behaviour metrics. Section \ref{sec:clusterResults} shows the results from that clustering and Section \ref{sec:discussion} discusses the implications of the results and future work. 

\subsection{Investment suitability} \label{sec:investmentSuitability}
Investors hire financial advisors who, in turn, recommend or distribute suitable financial products from investment dealers. The regulations for investment suitability for clients in Canada have been in place for decades and were formed through a collaboration of dealers, advisors, and regulators, with significant updates in 2009. This paper studies the KYC obligation that requires financial advisors and dealers to conduct due diligence on clients and take “reasonable steps” to establish such things as their identity, creditworthiness, investment needs, financial objectives, and risk tolerance. The KYC obligation is designed to protect clients and advisors from unnecessary financial risk that does not align with the needs of the client, and ensure advisors and dealers are acting in good faith. 

\subsection{Know your client}
To fulfill the KYC suitability requirement, advisors meet with clients to determine the client’s identity, investment needs, financial objectives and circumstances, and risk tolerance. Many, but not all, will use a formal questionnaire to help gather this information and score the risk tolerance\footnote{Questionnaires are not limited to these criteria since regulators do not require a specific questionnaire but to take “reasonable steps” to understand client needs.}. An effective KYC protocol collects two types of information: (1) objective demographic information (legal identity), and (2) subjective information, from the perception of the client and their financial advisor, on the client's investment needs, financial objectives, investment knowledge, appetite for risk and circumstances. For example, the questionnaire typically establishes the client's identity by their full name, social insurance number, date of birth, address, and phone number.  For investment needs, financial objectives and circumstances, they are asked about their income, net assets, living expenses, time horizon for the investment account, potential withdrawal of funds from the account over a year, how they would change their portfolio based on the market changes, how they set aside savings, plan for retirement, and make retirement savings plan contributions. To help determine risk tolerance, they are asked about investment knowledge, dependants, debt, willingness to take on risk-based on situational questions, and what they want to accomplish with their wealth. 

Research in the area of effective KYC protocols is at the emergent stage and has focused on the collection and evaluation of KYC information. The main focuses of research by the financial community have been on the objective information for improving compliance to prevent illegal or terrorist activities and decreasing the cost associated with increased compliance. Where KYC research exists, it tends to focus on cost efficiency-distributed ledger systems \citep{moyano17}, how the financial crisis in the USA from 2007 to 2009 may have been affected due to non-compliance to US KYC regulations \citep{bilali11}, on using KYC to protect client accounts \citep{mondal16}, and on improving auditor effectiveness in evaluating KYC compliance \citep{smet11}. 

In contrast, few studies have been conducted to study the subjective information of the KYC obligation and their relationship to advisor and client investment behaviours, client investment objectives and outcomes, and dealer strategies to assist their advisors \citep{osc2015}. \cite{palma11} reviewed a number of existing risk tolerance assessment tools and concluded that while the neoclassical economic concept of risk tolerance is clear, its measurement through surveys is unclear. Since the economic definition of risk tolerance is a variation in future spending, many economists use questions that measure income volatility over time in order to assess risk tolerance. These questions are theoretically correct, but their performance as predictors of actual investment behaviour during volatile stock markets is mediocre \citep{guillemette12}.  

\subsection{Trading behaviour}
At the onset, the hypothesis for our research was that a thorough and complete assessment of an investor's KYC data should lead to an accurate determination of their risk tolerance and suitability requirements. In turn, those determinations should manifest downstream in trading behaviour and, eventually, in product recommendations, portfolio construction and investment outcomes. 

In this paper, we look to better understand the relationship between collected KYC information and trading behaviours through applications of behavioural finance and statistical analysis. Behavioural finance is the intersection of psychology and finance to explain the trends and actions of financial markets, institutions, advisors, and individual investors.  Behavioural finance has three main areas of application: analysis of patterns in stock returns, studying trading activity, and corporate finance \citep{sub08}. Our analysis focuses on trading activity. Our dataset encompasses over 23,000 clients who work with financial advisors at an anonymous investment dealer under the auspice of the Investment Industry Regulatory Organization of Canada (IIROC) regulatory regime. We use an extended RFM behavioural finance model \citep{lumsden08}. RFM models are used primarily in direct marketing to analyze customer behaviours through the recency of their last purchase, the frequency of their purchases, and how much is spent on each purchase. RFM models have been embedded in data mining algorithms \citep{birant11}. 

It is important to acknowledge that investor behaviour is a complex and dynamic topic. Investor behaviour is not only driven by the investor’s personal motives such as their goals and financial needs but it is also influenced by the advisor relationship, dealer processes, regulatory obligations, and market influences. While the dataset and analysis used in this paper are unique, we are not privy to some of the subjective or undocumented influences and we cannot include them in our algorithms. It is therefore impossible for us to determine why the KYC process is not leading to the outcomes we would expect. Our analysis has inspired the question “Could protocols be improved?” but we can’t answer the question without further research - which we discuss in Section 5. 

\section{Data description and feature engineering for behavioural finance}\label{sec:featureEngine}
The data for this analysis is provided by a registered investment dealer that has provided investment products and technology to Canadian retail investors for over 30 years. The dealer hitherto has approximately 200 advisors who work with approximately $23,000$ clients across Canada with over \$5 billion Canadian dollars (CAD) in assets. Clients typically have multiple accounts each with different purposes. For example, a client may have accounts for: (i) retirement savings; (ii) children's education savings; and (iii) other savings. In total, clients with advisors who work with the dealer have over $50,000$ accounts. They provide a variety of financial products and services designed to support independent advisors. Their focus is to provide positive outcomes to clients and advisors, and not to push certain financial products.

In this section, we describe the KYC information and trades and transactions recorded in the data. We use descriptive analysis to demonstrate the demographics of our data and that the data is of good quality. We describe the features engineered from the data to be used in clustering, including unique metrics that measure client behaviours. 
  
\subsection{Data description and processing}
The data is comprised of $52,025$ accounts for $23,970$ clients with associated KYC information, trade and transaction details from August 13th 2018 to August 12th 2019. The datasets were edited by the data donor prior to our receipt to ensure all client identifiers were anonymized consistent with Canada's Personal Information Protection and Electronic Documents Act (PIPEDA) and standard research ethics protocols. Even using anonymization practices, there is still the possibility that clients  could be identified using machine learning algorithms \citep{rocher19}. Therefore, no individuals will be identified or referenced in this paper and any subset of the data cannot be shared with readers. 

The data is organized into linked datasets where entries were uniquely determined by an anonymized account ID or other relational database information. The specific datasets we used are a KYC information dataset and a trades and transactions dataset. We created new features derived from both datasets that effectively supplement the KYC information with metrics that measure trading behaviours.

The data was processed by cleaning the data for improper entries (e.g., recording typos), transforming values into categories (e.g., grouping occupations into classifications), removing irrelevant, anonymized (e.g., contact information), or repeated (e.g., postal code in place of residence region) data. Any variable containing over 10 percent missing values or errors (e.g., `*' or `unknown') is removed to avoid excessive bias from imputation in our analysis. On the remaining data, imputation is conducted for each numeric and categorical feature based on existing values. For example, missing values in categorical variables such as `residency' are filled with mode value `Ontario' since more than 67\% of clients are from Ontario; missing values in numerical variables such as `annual income' are filled with mean income based on the job categories from KYC. See Table \ref{tbl:imputation} in Appendix B for more details on missing data.

\begin{figure}[!htbp]
\begin{centering}
  \includegraphics[width=10cm, keepaspectratio]{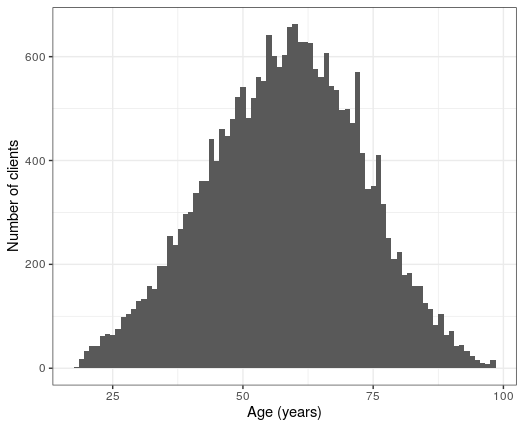}
  \caption{Distribution of client ages, where each bin contains one year.}
  \label{fig:allAges}
  \end{centering}
\end{figure}

Table \ref{tbl:clientDetails} shows the details of the pertinent objective KYC information. The distribution of client age is shown in Figure \ref{fig:allAges}. The client age distribution is unimodal, centred at 58.1 years, has a standard deviation of 14.1 years, and is slightly left-skewed. The minimum age is 18 years--the legal age to open an account in Canada--and the maximum is 98.
\begin{table}[!htbp]
\centering
\caption{Details of variables from clients' KYC information} 
\label{tbl:clientDetails}
\begin{tabular}{p{2cm}|p{5cm}|p{3cm}|p{2cm}}
Variable           & Summary     & Data type & Example values \\ \hline
Age                & Ages range from 18 to 98 years old, with average at 57.4 years   & Continuous  & 31 years old \\ \hline
Gender             &  $50.5\%$ male and $49.5\%$ female  &  Indicator & $M,F$                 \\ \hline
Residency          &  Province or Country or Region, with $~70\%$ from Ontario & Categorical & ON, UK, USEast, $\ldots$ \\ \hline
Annual income      & Gross annual income in CAD & Continuous & Multiples of $100$ between $\$1,000$ and $\$220,000$ inclusive \\ \hline
Investment knowledge & The self-reported investment knowledge of poor (2\%), fair (44\%), good (37\%), or sophisticated (17\%) & Ordinal           & $1$, $2$, $3$, or $4$ \\ \hline
Number of accounts & Clients can have more than one account & Ordinal & $1$,$2$,$3$,$\ldots10$                 \\ \hline
Marital status     & 67\% married, 18\% single, 11\% unknown and 4\% divorced & Categorical & M,D,S, or \mbox{*}               \\ \hline
Retirement indicator & The client's retirement status & Indicator & Yes, No                \\         
\end{tabular}
\end{table}

\begin{figure}[!htbp]
\begin{centering}
  \includegraphics[width=12cm, keepaspectratio]{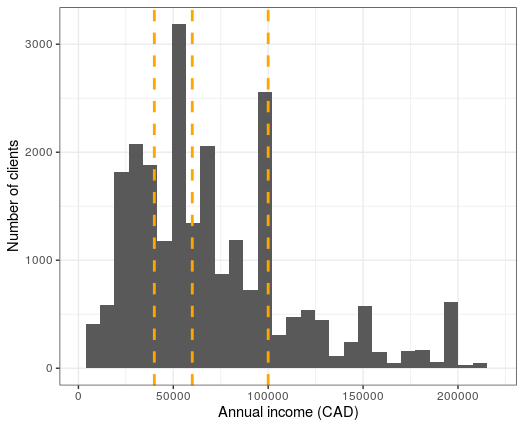}
  \caption{Distribution of client annual incomes. The vertical dotted lines represent the three quartiles at \$40k, \$60k, and \$100k.}
  \label{fig:annualIncomes}
  \end{centering}
\end{figure}

The distribution of account residency is shown in Table \ref{tbl:clientRes}, with the majority of accounts owned by clients in the province of Ontario. Figure \ref{fig:annualIncomes} shows the distribution of annual income. The income distribution has an average of \$$70,658$ and is right-skewed, with 50\% of clients making less than \$60k. There are also income spikes at \$50k and \$100k, \$150k and \$200k. Table \ref{tbl:numberOfAccountsUniqueClient} shows the number of accounts per client. Most clients have two accounts and few have five or more. 

\begin{table}[!htbp]
\caption{Distribution of residency for client accounts.}
\label{tbl:clientRes}
\centering
\begin{tabular}{rrrrrrrrrr}
  \hline
Location\footnotemark & ON & BC & AB & MB & NS & Other (CA) & Unknown & USA & UK \\ 
  \hline
Percentage & 65.19 & 14.63 & 12.00 & 3.94 & 2.59 & 0.92 & 0.41 & 0.26 & 0.06 \\ 
   \hline
\end{tabular}
\end{table}

\footnotetext{Ontario (ON), British Columbia (BC), Alberta (AB), Nova Scotia (NS), Canada (CA), United States of America (USA), United Kingdom (UK)}

\begin{table}[!htbp]
\centering
\caption{The number of clients by number of accounts.}
\label{tbl:numberOfAccountsUniqueClient}
\begin{tabular}{rrrrrrrrrrr}
  \hline
Unique accounts &   1 &   2 &   3 &   4 &   5 &   6 &   7 &   8 &   9 &  10 \\ \hline 
Number of clients & 5475 & 7659 & 6661 & 3051 & 775 & 222 &  79 &  40 &   4 &   4 \\ 
   \hline
\end{tabular}
\end{table}

Our dataset contains a combination of trades and transactions for each client. We reserve the word ``trades" for any interaction with mutual funds, stocks, securities, and bonds, and ``transactions" for any interaction that does not include those interactions such as collecting dividends and interest. Trades are logged as orders, which are either active, inactive, filled, rejected, cancelled, or expired. In this paper, only filled orders are studied and the study of investor behaviours through all of their order history and is deferred to future work. 

Each trade and transaction is recorded with the type of product or transaction, size, value, currency type, security identification code, order date, process date, value date, and more. Using the trades and transaction dataset, we determined the variables that we believe contain information on client behaviours and developed new metrics using feature engineering to measure client behaviour.

\subsection{Feature engineering}
Feature engineering in data science is the process of using industry knowledge about data to construct metrics or ``features" that can act as a measure for a quantity to be used in a machine learning model \citep{zheng18}. Features generated from an RFM model can be used in conjunction with a machine learning algorithm \citep{anitha19}. We construct features that using objective and subjective KYC information, and trade and transaction information that we believe to be related to client investment behaviour. Our features are an extension of an RFM model and fall into four categories: recency, frequency, monetary, and profile (RFMP).

The RFMP features are aggregated into a cross-sectional dataset that is static in time, where the cross-section is calculated on the last day recorded (August 12th 2019) in the dataset. Table \ref{tbl:features} lists the features used for the clustering algorithm described in Section \ref{sec:clustering} and to generate the results shown in Section \ref{sec:clusterResults}. We now describe each type.
\begin{table}[]
\caption{The RFMP features engineered from the dataset}
\begin{tabular}{l|p{6cm}|p{6cm}}
Feature type & Description & Variables \\ \hline
Recency & Number of days since last trade on record & Days between the most recent trade date and August 12, 2019 \\ \hline
Frequency & Total number of trades

\hfill \break Average number of days between trades & Number of trades between first trade date and August 12, 2019 \newline Number of days divided by number of trades since first trade day \\ \hline
Monetary & \begin{tabular}[c]{@{}p{5.5cm}@{}} Buy and sell size totals\\ Buy and sell size minimum and maximum\\Trade size by type\\ Variability of trade size by type\end{tabular} & \begin{tabular}[c]{@{}p{5.5cm}@{}}\underline{\it Third-party initiated trade type}\\ Dividends, income distribution, interest\\ \underline{\it Systematic trade type}\\ Auto-withdrawal, pre-authorized contribution, asset allocation, reinvested dividend \\ \\ \underline{\it Periodic trade type}\\ Buys, sells, contribution, exchange, payment, electronic funds transfer (EFT), withdrawal, EFT deposit, tax-free savings account (TFSA) contribution, spousal contribution, redeems\end{tabular} \\ \hline
Profile & KYC information

Financial descriptors (e.g. number of accounts)& Age, gender, residency, annual income, investment knowledge level, number of accounts, marital status, retirement indicator
\end{tabular}
\label{tbl:features}
\end{table}

Profile features describe the client as who they are and what their financial goals are. Commonly, they are considered influential factors to the behaviour of the client \citep{foerster17}. Profile features are generated from KYC and account information for each of the clients. Some of the profile features were immediately ready for usage (for example, the time horizon of the account) whereas other variables needed to be derived; age in years is calculated from birth dates and the number of accounts is determined by searching the database for client accounts. 

The recency feature is calculated as the number of days since a client's most recent trade or transaction. The frequency features are calculated through a client's overall amount of trading throughout the history of the dataset. These two features types provide some information on their own, but when used together are more than the sum of their parts. If they have a large total number of trades (frequency) and months since their last trade (recency), this means they have a ``burst" investing behaviour.  These feature types when used together provide an interesting picture of client behaviours. 

The monetary features are features engineered from trade and transaction amount details, rather than their temporal attributes. Specifically, a trade size multiplied by the value for each unit is the total monetary value in CAD, which we will refer to as the trade amount.  If we looked at each trade as equivalent--similar to recency and frequency--then we will incorrectly consider that purchasing a stock is the same as re-investing a dividend. The stock purchase is an active trade that a client or advisor initiates, whereas a re-invested dividend is not. We classify trade sizes into the three metrics given by 
\begin{align}
\begin{split}\label{eq:tradeBehave1}
    Third{\text -}party \ initiated \ trade \ size ={}& Dividend + Income \ distribution + Interest,
\end{split}\\
\begin{split}\label{eq:tradeBehave2}
    Systematic \ trade \ size ={}& Auto \ withdrawal + Pre {\text -} authorized \ contribution + \\
         & + Asset \ allocation + Reinvest \ dividend,
\end{split}\\
\begin{split}\label{eq:tradeBehave3}
    Periodic \ trade \ size ={}& Buy \ (securities) + Sell \ (securities) + Contribution + Exchange \\
        & + Payment + Electronic \ funds \ transfer \ (EFT) + Withdrawal \\
        & + EFT \ deposit + TFSA + Spousal \ contribution + Redeem
\end{split}
\end{align}
where the descriptions of the trade types can be found in Appendix A. Third-party initiated trades are comprised of trade types that are initiated by a third party, such as a coupon collected as cash from a bond. Systematic trades are comprised of self-imposed automatic investment strategies, such as an automatic monthly withdrawal from savings to purchase a mutual fund. Periodic trades are client or advisor initiated trades and transactions, such as an unscheduled purchase of a mutual fund for a TFSA. 

\begin{figure}[!h]
\begin{centering}
  \includegraphics[width=14cm, keepaspectratio]{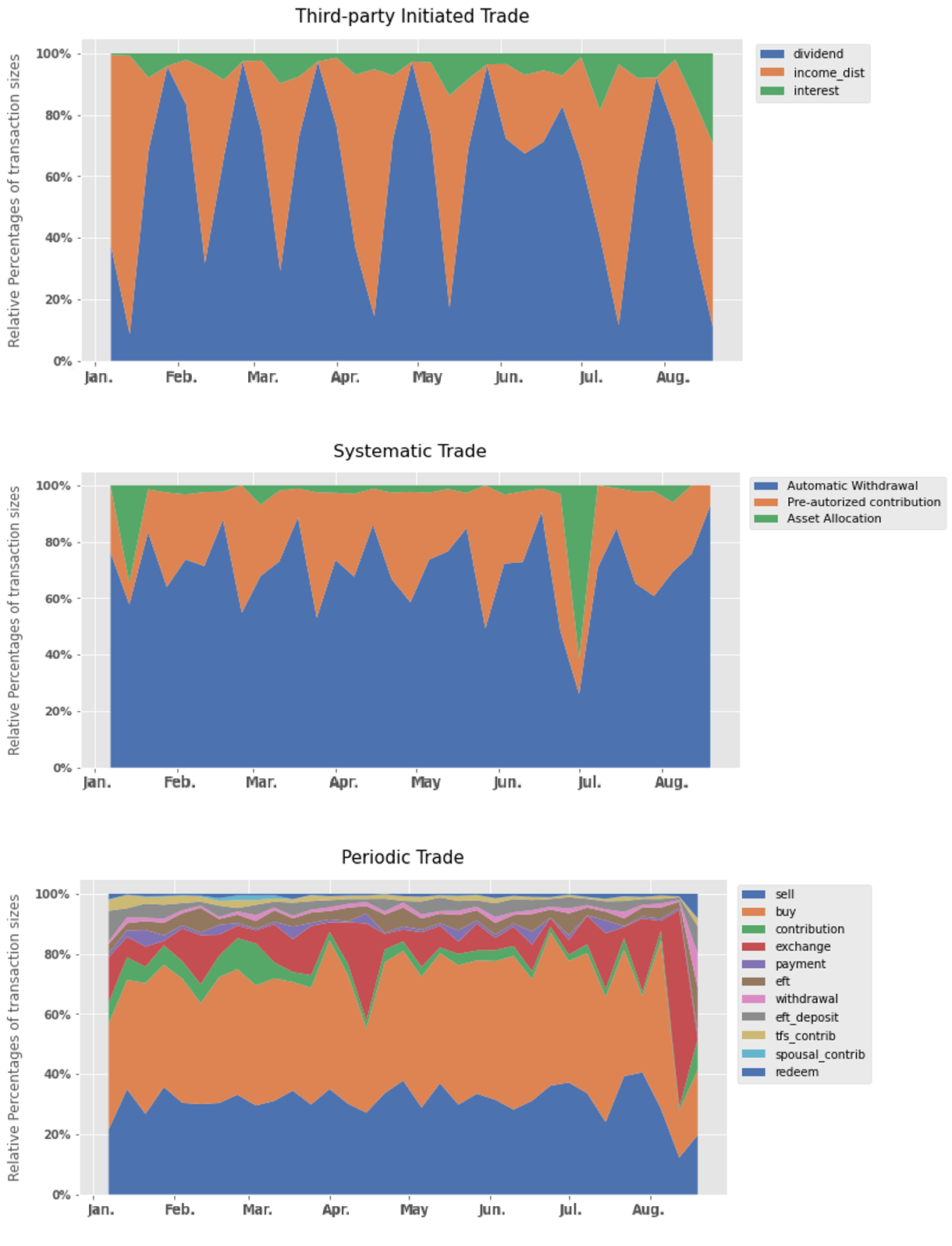}
  \caption{The relative percentage of transactions sizes from the three behavioural metrics versus time (January to August 2019). Top, middle, and bottom panels correspond to third-party initiated, systematic, and periodic trades, respectively.}
  \label{fig:pcnt}
  \end{centering}
\end{figure}

Figure \ref{fig:pcnt} shows the relative percentages of transaction sizes comprising the three behavioural metrics in Equations (\ref{eq:tradeBehave1}) to (\ref{eq:tradeBehave3}) versus time. For third-party initiated trade size, dividend and income distribution dominate most of the transactions, and there appears to be a cyclical trend for dividends paid at the beginning of every month. For systematic trades, automatic withdrawal represents the majority of the feature size and has an obvious cyclical trend. There are spikes for asset allocation at the beginning of the year and six months in; a bi-annual cycle for asset allocations in systematic trades. For the periodic trades, the buy and sell types dominate without any cyclical trends.

The features we engineer in this section are used directly as variables in our clustering model in Section \ref{sec:clusterResults}. The next step is to take our engineered features and use them in a clustering algorithm. The theoretical underpinnings for our algorithm are described in the next section, which is followed by empirical results from clustering in the subsequent section.

\section{Clustering theory and methods} \label{sec:clustering}

Clustering is an unsupervised machine learning algorithm that is used to draw inferences about grouping commonalities from like-individuals in high dimensional data. It is a popular method for exploratory data analysis that finds previously unknown structures in data without specifying the underlying data generating process. Clustering is a powerful technique used in many fields, such as identifying fake news \citep{hosseinimotlagh18}, bioinformatics \citep{krishna99,lan18}, text mining \citep{berry04}, and wireless sensor networks \citep{abbasi07}. 

Clustering bears the task of grouping our set of clients by considering the similarity of their attributes and trading behaviour \citep{xu08}. For obvious reasons, we are interested in applications of clustering for financial data analytics \citep{khac12}, particularly the area of Behaviour Clustering Analysis (BCA). Popular clustering algorithms used in this field are $k$-means \citep{steinley06} and $k$-modes \citep{huang98,chaturvedi01,huang03}. In this section, we introduce the $k$-prototypes algorithm that allows for both continuous and categorical data to cluster clients based on their similarity. Next, we introduce $t$-distributed stochastic embeddings that reduces the dimensions of the data based on the similarity of each data point. The embeddings display the data in low-dimensions by similarity, while the clustering algorithm identifies the clusters among the data points. 

\subsection{\texorpdfstring{$k$}{k}-prototypes clustering}
The $k$-prototypes algorithm used here is similar to the $k$-means algorithm, where $k$-prototypes incorporates methods for including categorical data \citep{Huang97clusteringlarge}. Suppose we have a set of $N$ accounts each with a unique identifier or index in the set $\mathcal{N}=\{1,2,\ldots,N\}$. The goal of any clustering algorithm is to put clients into $k$ groups or clusters such that
\begin{itemize}
    \item each client is put into exactly one cluster;
    \item clients within a cluster have similar attributes; and
    \item clients in different clusters have dissimilar attributes.
\end{itemize}
Mathematically, the $k$ clusters form a partition\footnotetext{A partition of any set $A$ is a set of subsets $A_1,A_2,\ldots,$ that are mutually disjoint ($A_i\cap A_j=\phi$ for all $i\neq j$) and exhaustive ($\cup_i A_i=A$).} of the the client index set into $k$ subsets. Let $\mathcal{N}_{\ell}$ denote the set of client indices for all clients in cluster $\ell$, $\ell=1,2,\ldots,k$, and $\mathcal{P}_\mathcal{N}=\{\mathcal{N}_1,\mathcal{N}_2,\ldots,\mathcal{N}_k\}$ denote the partition of the client index set. Furthermore, let $n_{\ell}$ denote the number of clients in cluster $\ell$, such that $\sum_{\ell=1}^kn_{\ell}=N$.

Each client has attributes that describe the individual given by their attribute vector $x_i,~i=1,\ldots,N$. These attributes are a combination of $p$ numeric variables (e.g., age) and $q$ categorical variables (e.g. marital status). Without loss of generality, we put the numeric attributes in the first $p$ positions of the attribute vector and the categorical attributes in the last $q$ positions giving
\begin{equation}
    x_i=(\underbrace{x_{i1},x_{i2},\ldots,x_{ip}}_{\mbox{numeric}},\underbrace{x_{i(p+1)},\ldots,x_{i(p+q)}}_{\mbox{categorical}}).
\end{equation}

The clustering algorithm works in an iterative fashion according to the following steps.
\begin{enumerate}
    \item Initialize the centroid (location) of the clusters by selecting $k$ clients as ``prototype" centroids.
    \item Allocate the clients to the clusters with the closest centroid. 
    \item Compute an overall cost of the allocation by computing total distance of all clients from their assigned centroids.
    \item Update cluster centroids.
    \item Re-allocate the clients to the clusters with the closest (updated) centroid.
    \item Compute the overall cost by computing total distance.
    \item Iterate steps 4-6 until there is no change in the overall cost and output the clusters.
\end{enumerate}

We kickoff the clustering party by randomly selecting $k$ clients to serve as the initial \emph{centroids} (locations) of the clusters. Specifically, the initial centroids are given by the attribute vectors of the randomly-chosen $k$ clients and are denoted by
\begin{equation}
    c_{\ell}=(\underbrace{c_{\ell 1},c_{\ell 2},\ldots,c_{\ell p}}_{\mbox{numeric}},\underbrace{c_{\ell (p+1)},\ldots,c_{\ell (p+q)}}_{\mbox{categorical}}),~\ell=1,\ldots,k,
\end{equation}
where $c_{\ell j}$ is the cluster-$\ell$, attribute-$j$ centroid. Attributes in the centroid vectors are positioned in exactly the same order as in the client attribute vectors. As we shall see, as clusters are formed the centroids get updated according to the individuals within each cluster.

After initializing the cluster centroids, we need some way of deciding how to put the clients into the clusters so that individuals within clusters are similar (close) and individuals across clusters are dissimilar (far apart).  To measure the similarity between client $i$ and cluster $\ell$ we use the distance metric
\begin{eqnarray}
    d(x_{i}, c_\ell) = \sum_{n=1}^{p} \sqrt{(x_{in} - c_{\ell n})^2} + \sum_{n=p+1}^{p+q} \delta(x_{in}, c_{\ell n}),
 \label{eq:k_proto_eq_1}
\end{eqnarray}
where
\begin{equation}
    \delta(a,b)=\left\{\begin{array}{ll}
    1     & \mbox{if }~a\neq b  \\
    0     & \mbox{for }~a=b
    \end{array}\right..
\end{equation}
Note that the distance metric is zero if and only if the attribute vector is exactly the same as the centroid and if there are no categorical variables ($q=0$) then $d(\cdot,\cdot)$ is the usual Euclidean distance.

For client $i$ the distance between its attribute vector and each of the $\ell$ cluster centroids are computed, $d(x_{i}, c_\ell), \ell=1,\ldots,k$, and the client is placed in the closest cluster (e.g., minimum distance).  This is done for all $N$ clients (the clients initially chosen as centroids will clearly be placed in the correct cluster), with each client assigned to exactly one of the $\ell$ clusters.

After all clients are assigned to a cluster, the overall distance between individuals and their cluster centroid is computed by the cost function
\begin{eqnarray}
    J = \sum_{\ell=1}^{k} \sum_{i\in\mathcal{N}_{\ell}}d(x_{i}, c_{\ell})
 \label{eq:k_proto_4_cost}
\end{eqnarray}
The cluster centroids are updated by independently finding the middle for each cluster's attributes. For the numeric attributes, the centroids are updated to be the within-cluster average value. Specifically, the updated $j$-th attribute for cluster $\ell$ is 
\begin{eqnarray}
    c_{\ell j} = \frac{1}{n_{\ell}}\sum_{i\in\mathcal{N}_{\ell}} x_{ij}, ~~j=1,\ldots,p.
 \label{eq:k_proto_eq_2}
\end{eqnarray}    
The categorical attributes of each cluster are updated using the mode, given by  
\begin{eqnarray}
c_{\ell j} = \mathbb{M}\left(x_{ij}|i\in\mathcal{N}_\ell\right)
 \label{eq:k_proto_eq_3}
\end{eqnarray} 
where $\mathbb{M}$ is the mode function. Next, we re-allocate each client to clusters using the minimum distance between the client attribute vector and the updated cluster centroids.  After re-allocation, the overall cost is computed using Equation \ref{eq:k_proto_4_cost}. If the total cost is unchanged from the previous iteration, we stop; otherwise, the cluster centroids are updated and clients are re-allocated. This is repeated until the total cost function is unchanged. 

Since the initial set of $k$ cluster centroids (e.g., $k$ clients serving as initial centroids) is chosen randomly, the clustering process is repeated for a large number of randomly-chosen initial cluster centroids to better search for the global minima of the cost function.  Each initial cluster centroid produces clusters and their total cost.  The best (and final) cluster is the one that minimizes the cost function over all randomly-chosen initial cluster centroids. Typically it is infeasible to look at all possible $k$ initial cluster centroids, which is the reason for the random sampling of the initial cluster centroids.  For example, with $N=25000$ clients and $k=5$ clusters, the number of possible ways of choosing the initial cluster centroids is $\frac{25000\times24999\times24998\times24997\times24996}{5!}$ which is an infeasible number of possibilities to examine.

\subsection{Visualizing clusters - \texorpdfstring{$t$}{t}-distributed stochastic neighbour embeddings}\label{sec:tsne}
Visualizing high-dimensional data by projecting it onto a lower-dimensional space is commonly used \citep{data_project_citation}. The computationally efficient dimensionality reduction tool used herein is the $t$-distributed stochastic neighbour embeddings ($t$-SNE) \citep{maaten08}. The $t$-SNE method provides a significant dimensionality reduction from high dimensional data to two- or three-dimensions while preserving the significant structure. This method is a nonlinear mapping which, as opposed to linear mappings, performs better for preserving the local structure of data--that is, this method keeps similar clients close together in a low-dimensional visualization. This is important for visualizing clusters since we are using a clustering method that evaluates clients by their similarity. Therefore, the $t$-SNE method creates a map of clients based on their similarity, and then we independently apply the clustering algorithm to the data--all without specifying the data generating process.  

\begin{figure}[!hbt]
\begin{centering}
  \includegraphics[width=16cm, keepaspectratio]{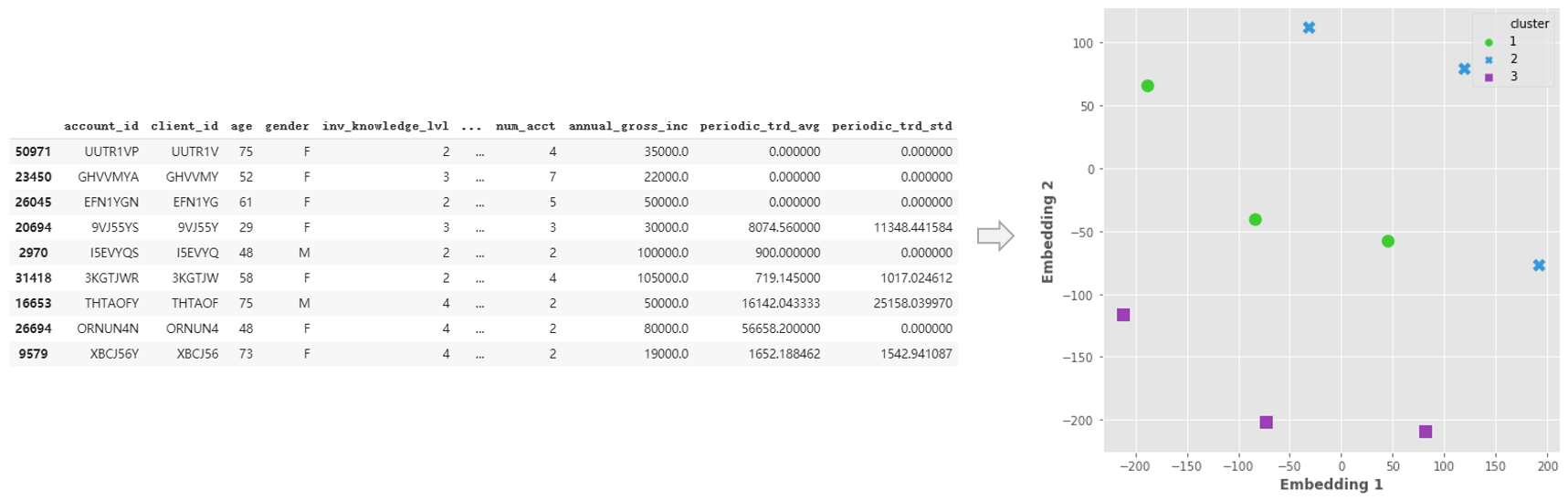}
  \caption{A $t$-SNE's 2-D projection for a small sample of client data.}
  \label{fig:tsne_demo}
  \end{centering}
\end{figure}

Figure \ref{fig:tsne_demo} displays the visualization of some sample client data; $t$-SNE is applied to project the high dimensional data into the 2-D space. 
For the $t$-SNE method, ``perplexity" is an important parameter that affects the visual behaviour of data projection. Different datasets require different perplexities to display the clustering--or lack thereof--features present in the data. According to \citep{maaten08}, the perplexity can be viewed as the algorithm's method to measure the number of effective nearest neighbours with typical values between 5 and 50. Choosing the perplexity value requires the user to tune it during the modelling process. There is no standard method for specifying the perplexity value.  Furthermore, larger datasets require a larger perplexity \citep{perplexity_details}. For our dataset, the perplexity value is set to 200 to get a stable embedded data plot.

\section{Results}\label{sec:clusterResults}
In this section, we discuss the results of applying the  method described in Section \ref{sec:clustering} to the client data discussed Section \ref{sec:featureEngine}. The data cleaning, feature engineering, clustering algorithm, $t$-SNE embedding visualization, and analysis are implemented using Python version 3.6 and \texttt{R} version 3.5.3 \citep{coreR}. The implementation of the $k$-prototypes clustering algorithm originated from a GitHub repository \citep{kmodes} and the $t$-SNE algorithm used for data visualization is in the \texttt{sklearn} Python package \citep{scikit-learn}. 

\begin{figure}[!htb]
\begin{centering}
  \includegraphics[width=12cm, keepaspectratio]{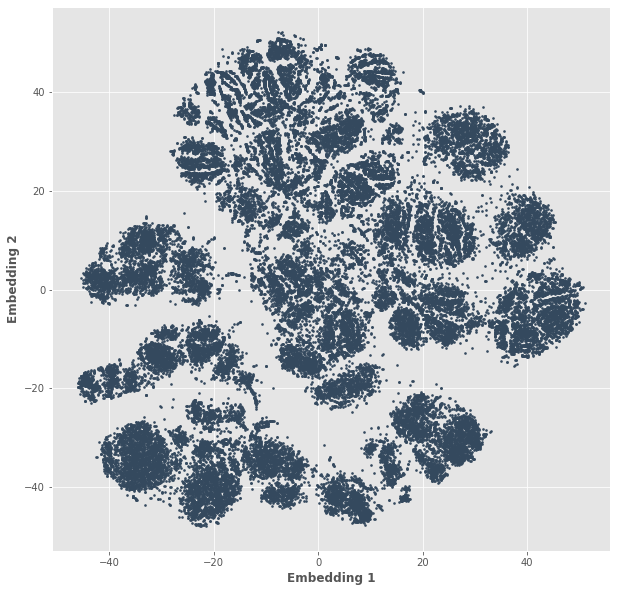}
  \caption{$t$-SNE visualization for the full data set projected onto two embeddings.}
  \label{fig:tsne_full_data}
  \end{centering}
\end{figure}

Figure \ref{fig:tsne_full_data} shows a two-dimensional similarity representation of the data using the $t$-SNE algorithm with a perplexity of 200\footnote{See Section \ref{sec:tsne} for discussion on perplexity for the $t$-SNE method}. Each point represents one client's attributes projected down to two dimensions, where the Euclidean distance between clients by their embedding represents a quantification of their similarity. The next step is to use the $k$-prototypes clustering algorithm to identify the optimal number of clusters $k$ for this client dataset.

\subsection{Choosing the optimal number of clusters}
\begin{figure}[!htb]
\begin{centering}
  \includegraphics[width=13cm, keepaspectratio]{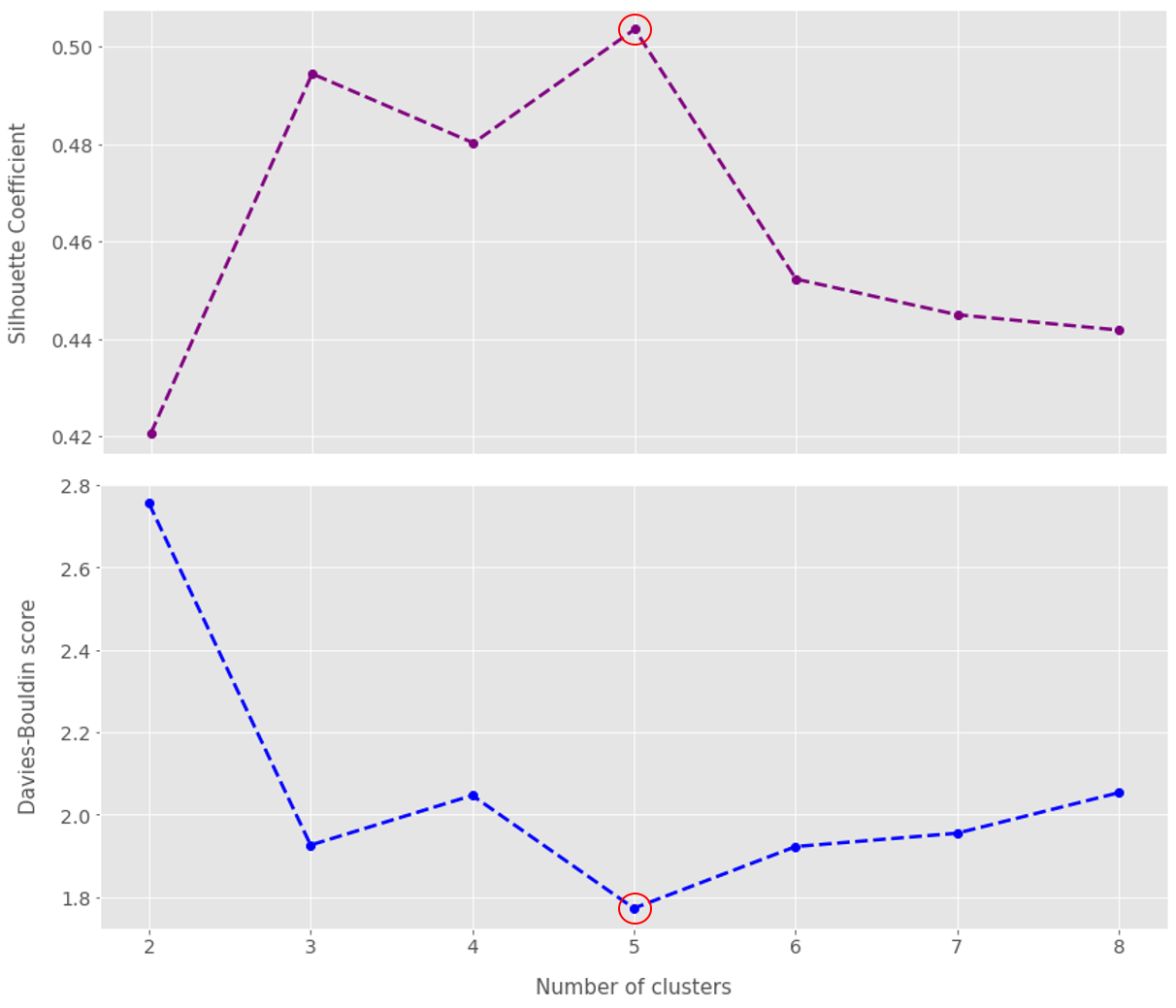}
  \caption{The top panel shows the average Silhouette coefficient and the bottom panel shows the DB score for different numbers of clusters. The optimal number of clusters is identified by the red circle at the elbow.}
  \label{fig:optimal_k}
  \end{centering}
\end{figure}

Two clustering performance evaluation methods are used to determine the optimal number of clusters: the Silhouette coefficient and the Davies-Bouldin (DB) score. The Silhouette coefficient \citep{silhouettes} compares the cluster membership classification of each client by comparing their similarity within and between clusters and indicates how well clients are assigned. The Silhouette coefficient of client $i$ in cluster $\mathcal{N}_\ell$ is defined as
\begin{eqnarray}
    S_i = \frac{b_{i} - a_{i}}{\max(a_{i}, b_{i})},
 \label{eq:silhouette_coefficient}
\end{eqnarray}
where $a_i$ is a similarity measure of client $i$ to clients within their cluster given by
\begin{eqnarray*}
a_i = \frac{1}{|\mathcal{N}_i|-1}\sum_{j\in\mathcal{N}_\ell,j\neq i}d(x_i,x_j),
\end{eqnarray*}
and $b_i$ is a similarity measure of client $i$ to the clients in the most similar or closest neighbouring cluster given by
\begin{eqnarray*}
b_i = \min_{g\in\{1,2,...,k\},g\neq\ell}\left[\frac{1}{|\mathcal{N}_g|}\sum_{j\in\mathcal{N}_g}d(x_i,x_j)\right].
\end{eqnarray*}
The best assignment value for the Silhouette coefficient is 1 and the worst value is -1, and values near 0 indicate overlapping clusters. Negative values generally indicate that a client may be poorly assigned, as a different cluster is more similar. Figure \ref{fig:optimal_k} shows average Silhouette coefficient $S=\frac{1}{N}\sum_{i=1}^NS_i$ for $k=2$ to $8$ clusters. The average Silhouette coefficient is maximized for this clustering method when we choose $k=5$ clusters.

The DB score \citep{db_index} is another cluster partition evaluation metric that compares the similarity between clusters with the size of the clusters themselves. The DB score is calculated as
\begin{eqnarray}
    DB = \frac{1}{k}\sum_{i=1}^{k}\max_{j \neq i} \left(\frac{s_{i} + s_{j}}{d_{ij}}\right)
 \label{eq:db_index}
\end{eqnarray}
where $k$ is the number of clusters, $s_{i}$ is the average distance of all clients in cluster $i$ from the centroid $c_i$, and $d_{ij}$ is the distance between cluster centroids $c_i$ and $c_j$. The DB index quantifies the density of clusters and clusters which are farther apart. Hence, the DB index decreases as separation between the clusters increases. Similarly to the averaged Silhouette coefficient, the second plot in Figure \ref{fig:optimal_k} indicates a $k=5$ clustering partition yields the optimal clustering results.

\begin{figure}[!htb]
\begin{centering}
  \includegraphics[width=12cm, keepaspectratio]{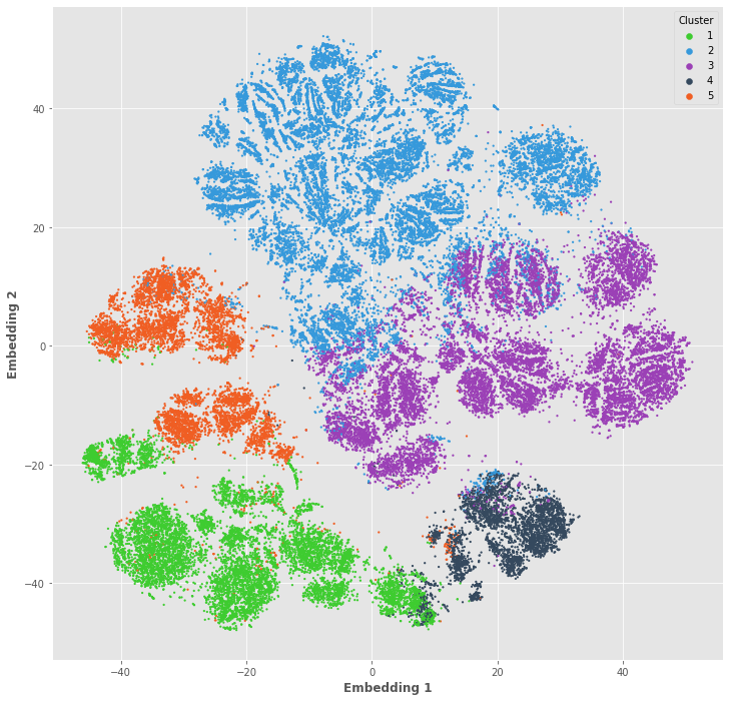}
  \caption{$t$-SNE visualization for the full dataset by cluster projected onto two embeddings.}
  \label{fig:clusterResult}
  \end{centering}
\end{figure}

Figure \ref{fig:clusterResult} shows the overlaid cluster membership on the $t$-SNE visualization. Among the 5 clusters, cluster 1 has 19\% of the clients and its data points are green on the embedding map, cluster 2 has the largest portion of clients with (36\%) and is labelled blue, cluster 3 has 27\% of clients and is labelled purple, cluster 4 the least portion (7\%) of clients and labelled black, and cluster 5 has 12\% of clients and is labelled orange. 

From the two-dimensional embedding map in Figure \ref{fig:clusterResult}, there are distinct boundaries between clusters 2, 3 and clusters 1, 4, 5. There are overlaps between clusters 1 and 5, clusters 2 and 3, and clusters 1 and 4. It is noteworthy that higher dimensional embedding can reveal other higher-order boundaries that distinguish these overlapped clusters. The projection from three-dimensions to these two dimensions creates the visual appearance of overlapping.

\subsection{Within cluster analysis}\label{sec:withinCluster}

\begin{figure}[!htbp]
\begin{centering}
  \includegraphics[angle=90,height=22cm, keepaspectratio]{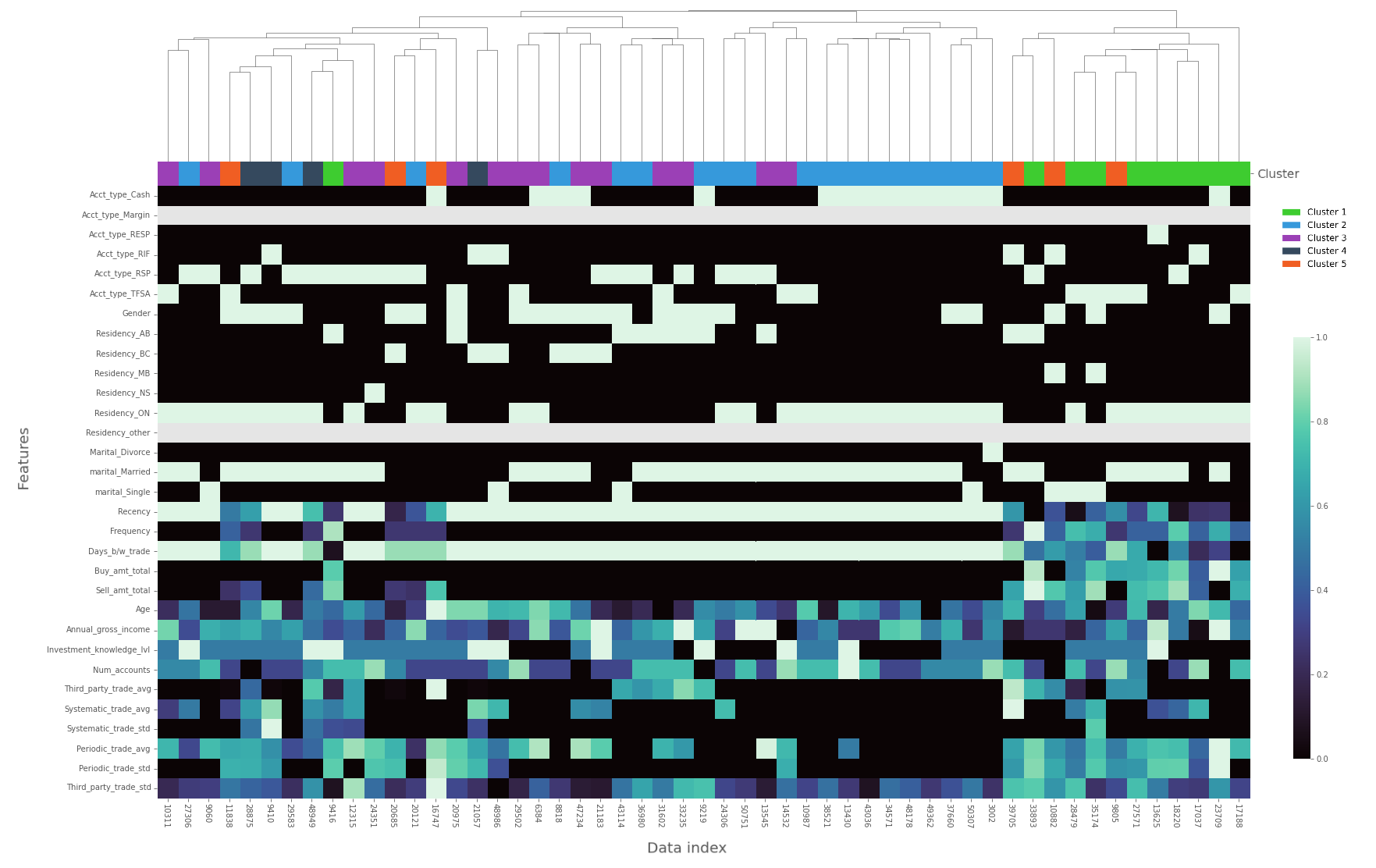}
\caption{A dendrogram of the clustering result with a heat map. Each attribute value is scaled to lie in the interval $[0,1]$, where the minimum attribute value is scaled to zero and maximum value scaled to one. Larger values (more white) indicate a larger relative value to other members in the same attribute.}
  \label{fig:clust_dendro}
  \end{centering}
\end{figure}

Figure \ref{fig:clust_dendro} shows a tree-structured dendrogram with a heat map to visualize the pattern within and between clusters' attributes. A sample of 53 clients from the dataset is selected by stratified random sampling, where each cluster represents a stratum and the relative number of selected individuals is proportional to the cluster size. Each row of the dendrogram shows an individual client's attributes, and the columns show the features used in clustering. The first column is the clustering labels from Figure \ref{fig:clusterResult}. For each remaining column, a heat map is presented with the scaled values using the range of each attribute. The minimum value of the attribute is scaled to zero (black) and the maximum value is scaled to 1 (white), and the rest of the values between the minimum and maximum are mapped on a linear scale. The dendrogram rows are ordered by distance between the clients' attributes using a hierarchical structure shown on the left side of the diagram.

Table \ref{tab:within-cluster} summarizes the mean values of the numeric features for each cluster. These mean values are the numeric attributes of the centroids (location) of the optimal clusters. Figure \ref{fig:clusterResult} and Table \ref{tab:within-cluster} demonstrate the following patterns between each of the clusters:
\begin{itemize}
    \item Clusters 1 (green) and 5 (orange) are similar in their demographics and trade types, but cluster 5 trades less often with smaller periodic trade sizes. 
    \item Cluster 2 (blue) is distinct from the others where they are largely inactive in their trading. 
    \item Clusters 3 (purple) and 4 (gray) are similar, except that cluster 3 makes larger, less frequent trades and cluster 4 utilizes larger systematic trades. 
\end{itemize}  
\setlength{\tabcolsep}{7pt}
\renewcommand{\arraystretch}{1.5}
\begin{table}[!htbp]
\centering
\caption{Mean values of the numeric features of the optimal cluster centroids for each cluster} \label{tab:within-cluster}
 \begin{tabular}{ c | >{\columncolor{YellowGreen}} c | >{\columncolor{CornflowerBlue!80}} c | >{\columncolor{Purple!70}} c | >{\columncolor{Black!30}} c | >{\columncolor{YellowOrange!70}} c} 

  Cluster & 1 & 2 & 3 & 4 & 5 \\  \hline
 Age (years) & 58.7 & 55.5 & 59.6 & 64.5 & 57.9 \\  \hline
 Annual gross income (CAD) & 72310.11 & 72623.69 & 69397.60 & 62229.89 & 69955.47 \\  \hline
 Investment knowledge level  & 2.69 & 2.70 & 2.68 & 2.84 & 2.70 \\
 \hline
 Number of accounts & 3.07 & 3.03 & 3.05 & 2.85 & 2.89 \\
 \hline
 Recency (days) & 57.9 & 179.59 & 179.9 & 153.8 & 61.9 \\
 \hline
 Frequency (trades per day) & 5.77 & 0.006 & 0.0004 & 0.46 & 1.32 \\
 \hline
 Days between trades & 5.15 & 179.46 & 179.9 & 151.93 & 85.18 \\
 \hline
 Mean third-party trade (CAD) & 98.01 & 17.19 & 102.21 & 63.40 & 109.07 \\
 \hline
 SD third-party trade (CAD) & 79.13 & 7.51 & 57.69 & 46.17 & 57.23 \\
 \hline
 Mean systematic trade (CAD) & 350.08 & 22.34 & 292.90 & 946.09 & 251.61 \\
 \hline
 SD Systematic trade (CAD) & 25.53 & 0.13 & 0.11 & 671.11 & 0.35 \\
 \hline
 Mean periodic trade (CAD) & 36064.08 & 72.09 & 22071.42 & 11543.26 & 14060.87 \\
 \hline
 SD periodic trade (CAD) & 27685.31 & 0.71 & 12190.73 & 16335.76 & 12828.52 \\
\end{tabular}
\end{table}

\begin{figure}[!htbp]
\begin{centering}
  \includegraphics[width=16.5cm, keepaspectratio]{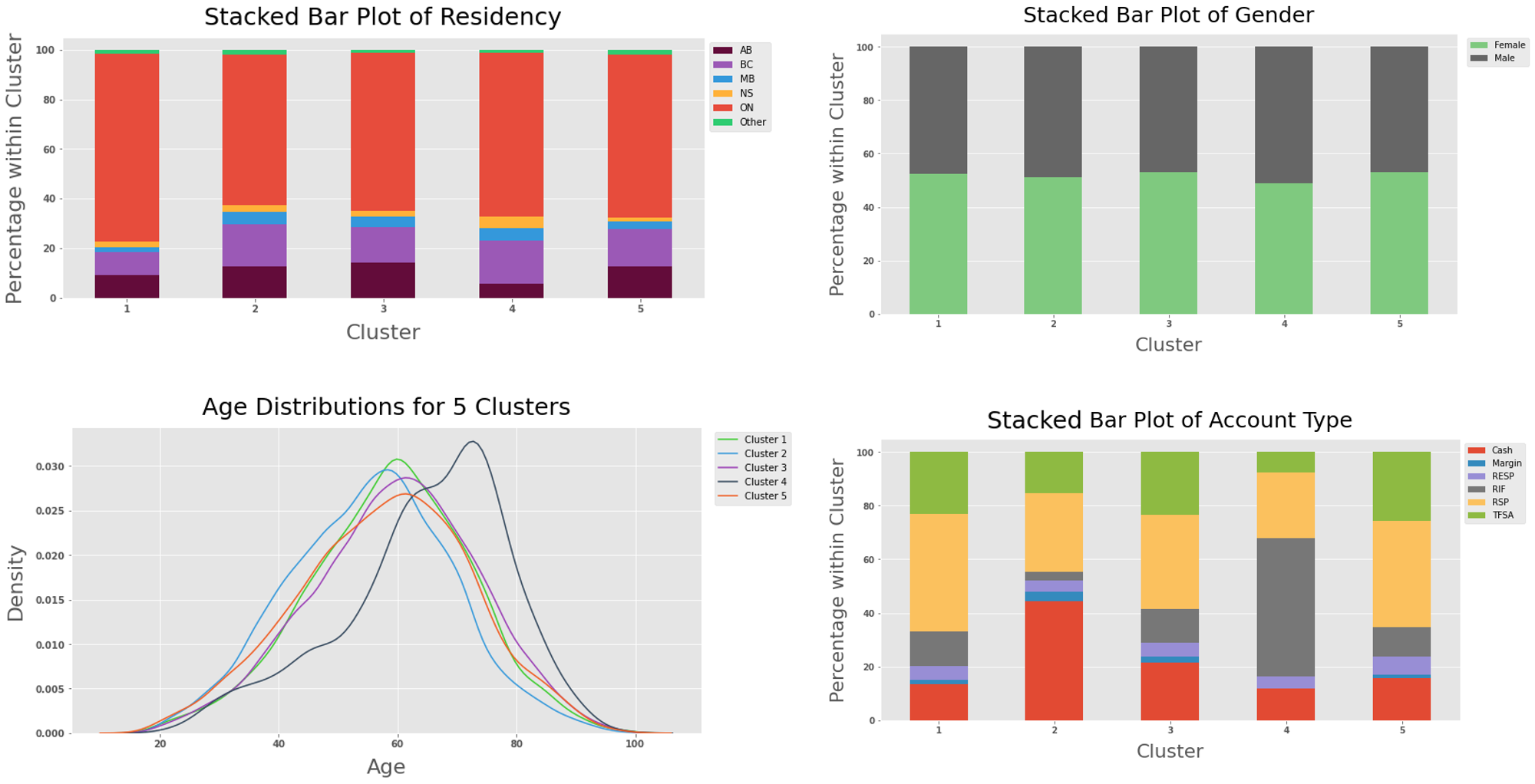}
\caption{Categorical and numerical distributions of clusters. Top left panel shows the residency distributions, top right shows the gender distributions, bottom left shows the age distributions, and bottom right shows the account type distributions for each cluster.}
\label{fig:clust_gender_resi}
  \end{centering}
\end{figure}

Figure \ref{fig:clust_gender_resi} shows the clustering results for categorical features. For the residency and gender features, there are no obvious differences between clusters. For the age feature, cluster 4 a high average age, and the distribution is left-skewed and appears almost bimodal. Clusters 1, 3 and 5 have similar age distributions. The cluster 2 age distribution appears shifted left and has younger clients compared to other clusters. The bottom right panel shows the percentages of the six account types in different clusters. Clients in clusters 1, 3 and 5 have similar account proportions. Cluster 2 has more cash accounts and cluster 4 has more RIF accounts.

\begin{figure}[!htbp]
\begin{centering}
  \includegraphics[width=12.5cm, keepaspectratio]{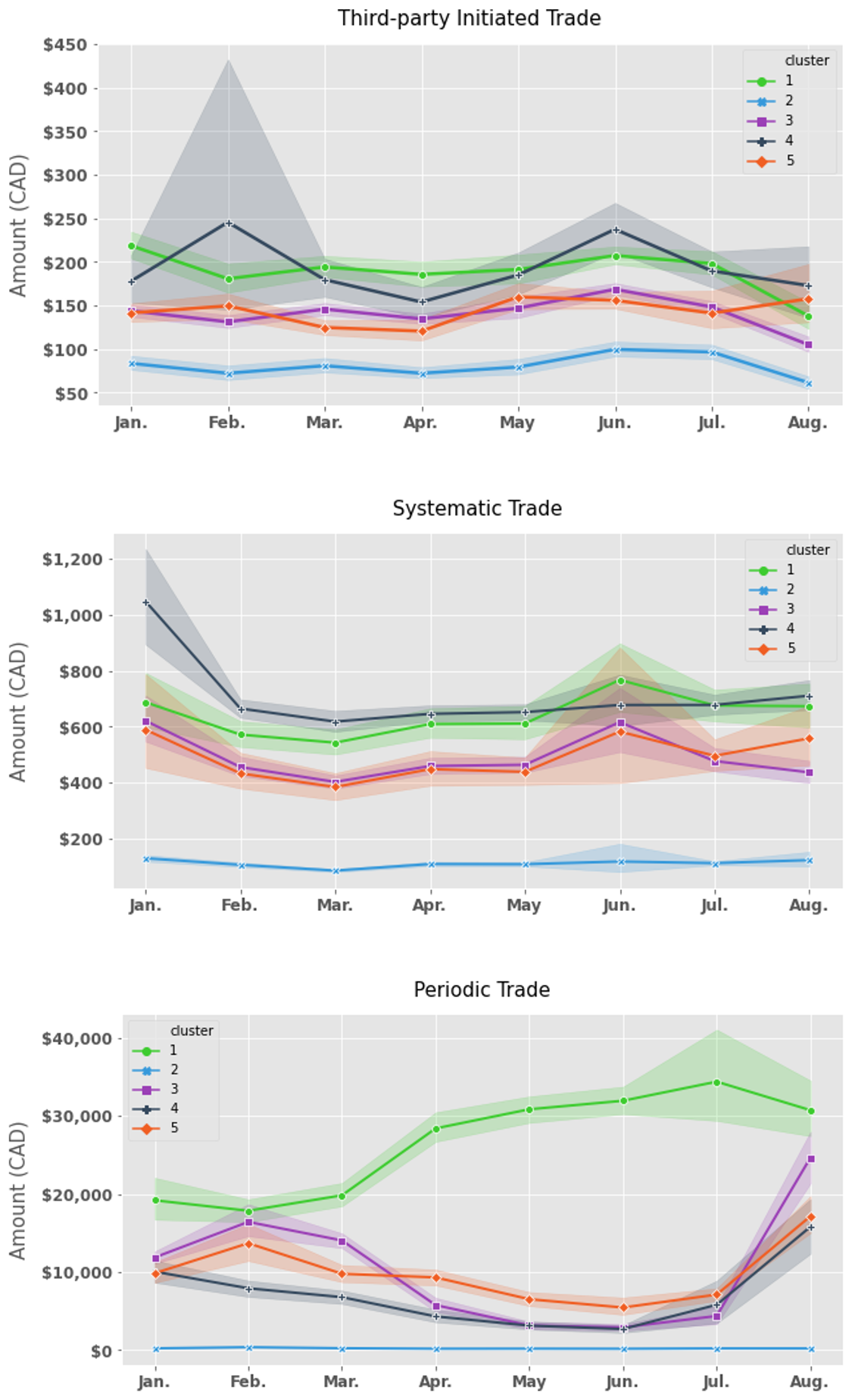}
  \caption{Cluster average trading amounts with 95\% bootstrapped confidence intervals versus time. Top, middle, and bottom panels correspond to third-party initiated trades, systematic trades, and periodic trades, respectively.}
  \label{fig:clusterResult2}
  \end{centering}
\end{figure}

Figure \ref{fig:clusterResult2} shows the monthly average trade amount over time, where the shaded areas are 95\% bootstrapped pointwise confidence intervals.  We note first the scale of each type of trade in the figure, where there are three different orders of magnitude. This may be caused by the nature of the trade types or by the number of elementary trade types within each of the trade type classes defined in Equations (\ref{eq:tradeBehave1}) to (\ref{eq:tradeBehave3}). 
\begin{itemize}
    \item For third-party initiated trades, cluster 4 has a relatively high trade amount and the largest volatility. Cluster 1 has similarly high trade amounts but less volatility. Clusters 3 and 5 have very similar trade amounts and volatilities that are smaller on average than the trade amounts and volatilities of clusters 1 and 4. Cluster 2 has the lowest average trade size and volatility.

    \item For systematic trades, a similar pattern to third-party initiated trades is reflected. Clusters 1 and 4 are again similar in the trade amount and volatility, with cluster 4 having slightly larger amounts on average except in June. Clusters 3 and 5 have almost identical average trade amounts except in August, and cluster 2 has the smallest average trade amount. An interesting aspect of all clusters is the peaks for the average trade amount evident in January and June.

    \item Cluster 1 dominates the periodic trade amounts, while cluster 2 has almost zero periodic trade amounts on average with very little volatility. Clusters 3 to 5 have similar trade amounts and volatilities, except in February and March when there is a slight peak before trending down for clusters 3 and 5. Clusters 3 to 5 all have an uptick in the average trade amount in July. There is a clear scale difference compared to the previous two trade types.
\end{itemize}

\begin{figure}[!hbt]
\begin{centering}
   \includegraphics[width=17cm, keepaspectratio]{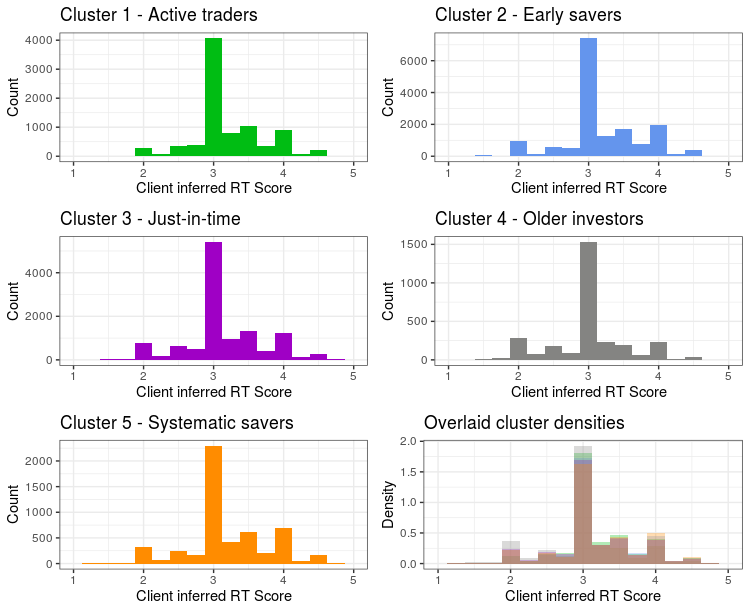}
  \caption{Inferred RT score distributions by cluster. The top four and bottom left panels are each cluster's distribution of the number of clients by inferred RT score. The bottom right panel is each of the clusters' risk score density overlaid.}
  \label{fig:riskToleranceScoresByCluster}
  \end{centering}
\end{figure}

Figure \ref{fig:riskToleranceScoresByCluster} shows the inferred risk tolerance (RT) score distributions for clients of each cluster. The majority of clients in each cluster's distribution (top four and bottom left panels) have a RT score close to three. Furthermore, each distribution appears quite similar, with smaller upticks at RT scores of two and four. The panel in the bottom right shows the overlaid translucent densities of each cluster, where the reddish-brown area is the shape that all clusters share. 

We investigated the similarity of these distributions using a parametric ANOVA comparison of client RT score means and a nonparametric Kruskal-Wallis test comparison of means \citep{kruskal52,mckight10}, for which both tests' null hypothesis were rejected with $P$-values $\leq2\times10^{-16}$ and 3.23$\times10^{-79}$, respectively. A post hoc analysis of a comparison of individual groups with adjusted $P$-values for multiple comparisons was conducted using Tukey's test \citep{tukey49} for ANOVA and the nonparametric Dunn's test \citep{dunn64} for Kruskal-Wallis test. The results of these tests are shown in Appendix C. These results suggest that clusters 3 and 4 have significantly different distributions from the rest. We investigated the difference in the distributions using the histogram density estimators (Figure \ref{fig:riskToleranceScoresByCluster}) in a a pairwise symmetric Kullback-Liebler (KL) plug-in estimator \citep{kullback51,ramirez04,wang05}. The KL estimator shows that the difference between the unlike-clusters' divergences (3,4) is not much larger than the like-clusters (1,2,5) divergences. The results of the symmetric KL estimators are shown in Appendix C.

From these analyses between the clusters in terms of the distribution of inferred RT scores, we can conclude that the distributions are similar, although there exists a statistically significant difference between the distributions. A smaller sample of points from each distribution would have a difficult time rejecting the null hypotheses of an analysis of variance test. The mean pattern and shape of risk tolerance distributions do not line up with what we would have expected. Clusters 1 and 4 are the most striking. Cluster 4 is demographically skewed towards older investors and we would expect to see RT scores weighted towards scores 1.0, 2.0 or 3.0. There are, in fact, only 15.7\% of clients in Cluster 4 who have less than a 3.0 RT score. Behaviourally, cluster 1 appears to pursue a riskier trading strategy and we would, therefore, have expected to see a strong weighting towards observations in the 4.0 to 5.0 RT score range. In fact, 14.8\% of cluster 1 clients fall into the 4.0 to 5.0 RT score range.

\subsection{From data to people -- Personas}
The cluster memberships are determined by the similarity of individuals, and we are interested in studying how the groups differ from each other. Using the plots and information presented heretofore, we summarize how the clusters differ using the most important variables to their cluster classification. We note that individuals from two different groups may appear similar, but they are classified based on subtle differences determined by the clustering algorithm. 

Using our understanding of investors and finance, we have created `personas' for clients to ease discussions and help understand the groups as real people and not just data. The five personas are as follows:
\begin{itemize}
\item Cluster 1: Active Traders (19\% of investors) trade frequently (weekly and monthly) and in large amounts. The pattern of trades is seemingly random and initiated manually. These investors had investments across a spectrum of accounts (mainly registered savings plans (RSPs) and TFSAs), and were of an ``average" age distribution and demographic. They had a derived risk tolerance rating that averaged 3.19 with standard deviation 0.63, where 1 is a low or preservative risk tolerance and 5 is high or aggressive. 
\item Cluster 2: Early Savers (36\%) never actively trade and instead rely on systematic transactions (auto-withdrawal, pre-authorized contribution, asset allocations). This group tended to  have investments in cash accounts and to be younger. They had a derived risk tolerance rating that averaged 3.18 with standard deviation 0.75.
\item Cluster 3: Just-In-time (27\%) initiate trades manually but far less frequently than Cluster 1 and in smaller amounts. These investors had investments across a spectrum of accounts (RSPs, TFSAs etc.), and were of an ``average" age and demographic. they had a derived risk tolerance rating that averaged 3.12 with standard deviation 0.73.
\item Cluster 4: Older Investors (7\%) trade infrequently and the trades were either initiated systematically or from a third-party (pre-authorized withdrawals, dividends and other disbursements). This cluster had an above average concentration of RIFs, and tended to be older. They had a derived risk tolerance rating that averaged 2.95 with standard deviation 0.71.
\item Cluster 5: Systematic Savers (12\%) trade recurrently (every 60, 90, or 120 days), in small amounts driven by systematic processes (dollar cost averaging) and periodic trading. These investors had investments across a spectrum of accounts (RSPs, TFSAs etc.), and of an ``average" age and demographics. They had a derived risk tolerance rating that averaged 3.19 with standard deviation 0.76.
\end{itemize}

\section{Discussion and Future Plans}\label{sec:discussion}
We have conducted a variety of approaches to analyze the client dataset to extract financial behaviours. We have constructed data summaries and extracted features that we believe capture financial behaviours, and included those summaries and features in a descriptive analysis. The features engineered from our data will directly affect the performance of future predictive models we are developing. We conducted a $k$-prototypes clustering algorithm on extracted features, where the cluster memberships were determined by minimizing a similarity cost function. We evaluated our clustering method using a Silhouette coefficient and a DB score, and analyzed the clustering results using the centroids generated by the algorithm and $t$-SNE visualizations. 

The ultimate goal of our research is to provide enhanced advice to clients and their advisors using both traditional and digital approaches.  The projects described herein are a path to attain that goal, providing the necessary algorithms to give information and advice in good faith. The projects not only support digital advice, but the results can be used to report to regulatory committees on how data-driven results can aid regulators in promoting financial wellness policies.

Moving forward, we will examine the behaviours of the clusters against the suitability and KYC protocols noted in this paper and then attempt to determine if those behaviours have a constructive or destructive impact on client outcomes. We also plan to examine the impact that advisor behaviours have on the analysis noted above while looking for evidence for whether we can change or nudge any or all of the noted behaviours. Previous research has determined that traditional characteristics explain only 12 percent of an investor's portfolio allocations \citep{foerster14,grace14,foerster17,linnainmaa18}. Our goal is to use new, sophisticated technologies to help examine the remaining 88 percent of unexplained investor behaviour \citep{grace19}. 

\subsubsection*{Trade and Asset Mix}
At the root of modern portfolio theory is the assumption that portfolio asset mix drives the portfolio’s inherent risk. The determination of suitability, based on the KYC, extends through portfolio construction to ensure that the portfolio’s asset mix is consistent with the investors risk tolerance. In our next phase of the project, we will use the same statistical techniques and dataset  above to examine whether the trading behaviour identified in each cluster is "suitable"--as defined by the prescribed regulations. We will complete this analysis by looking at the “asset mix” exhibited by each cluster. We will evaluate the security risk in the context of the client risk derived from the attributes of the cluster analysis. We will use security risk ratings (SRR) that are defined by industry  for each of the securities bought and sold and held by the client. These risk ratings are required by regulators under the Know Your Product protocols \citep{oscReport2019}. We will examine the trading behaviour and trade mix at specific points in time and then along a longitudinal continuum to see if the relationship changes over time. From this analysis, we will be able to determine if investor behaviour is ‘suitable’. We will examine how the trading behaviour exhibited by each cluster impacts their portfolios and the probability of achieving their desired outcomes. We will also look for evidence of whether the investor’s trading behaviour leads to unintended changes in the portfolio’s asset mix and risk characteristics over time. 

\subsubsection*{Portfolio Returns}
Where the analysis noted in the previous projects examine risk and the probability of success, we also plan to examine returns.  We will analyze the assumption that higher risk should lead to higher returns (in the long run) and presumably faster portfolio growth . Likewise, lower risk will presumably lead to more modest returns and preservation of capital. During this examination, we will use multiple methods to calculate returns including industry best practices  and regulatory guidance. 

\subsubsection*{Advice}
This project recognizes that investor behaviour is a complex event with a number of variables influencing behaviour. Spouses, family, friends, media and events, for example, can all influence the timing, characteristics and trajectory of behaviour. However, it is widely acknowledged that the investment advisor acts as the gate keeper for most investment trades and therefore, presumably, the trading behaviour \citep{marsden11,montmarquette12,ific12,kinniry14}. In this project, we will look for evidence to see if the advisor’s behaviour is influencing trading behaviour consistent with the KYC and suitability requirements. 

\subsubsection*{Investor Outcome Improvements}
In this project, we will take advantage of a second unique data set to examine whether it is possible to change or influence investor behaviours through new, systematic technologies. Using the same methodologies above, and the same set of investors, we will examine investor behaviour before and after a significant system enhancement implemented in November 2019 - leading into the market events of March 2020. We will make use of control charts to help determine the key variables that drive ‘risky’ behaviour over time. We will use this analysis will help assess the viability of potential new algorithms in the digital advice space. 

\newpage

\bibliography{references}

\newpage

\section*{Appendix A - Trade type descriptions}\label{appendix} 

\begin{table}[htbp!]
\centering
\caption{Types of trades in the client database}
\label{tbl:tradeTypes}
\begin{tabular}{p{2cm}|p{3.5cm}|p{8cm}}
Type & Examples & Description \\  \hline
Third-party initiated &  Dividend \newline Income \newline Distribution Interest & Third-party transactions are generated by product manufacturers and vary by product type – securities, ETFs, mutual funds, fixed income etc. The generation of these transactions does not require the participation of the advisor or investor and flow from the manufacturer to the dealer and then to the investor’s account. \\ \hline
Systematic & Auto Withdrawal \newline Pre-authorized Contribution \newline Asset Allocation \newline Reinvest Dividend & Systematic transactions are created by the advisor or investor to automatically generate on a prescribed timetable (for example monthly or quarterly). When these transactions are set-up, they can run for months or years without change until such time as the advisor or investor determine a revision is required because of new circumstances. \\ \hline
Periodic & Buy (securities) \newline Sell(securities) \newline Contribution \newline Exchange \newline Payment \newline Periodic \newline EFT Withdrawal \newline EFT deposit \newline TFSA \newline Spousal contribution \newline Redeem & Periodic transactions are initiated by the advisor or investor without a prescribed transaction amount or time frame. The description for these transactions can vary by product type – for example ``sell" refers to the disposition of a security while ``redeem" refers to the disposition of a mutual fund.  
\end{tabular}
\end{table}

\newpage

\section*{Appendix B - Imputation}\label{appendixB}

The details of specific variables that were imputed are shown in Table \ref{tbl:imputation}. We investigated each variable removed values by imputing the missing values and including them in the clustering algorithm. The clients with categorical variables that were between 5\% and 10\% missing were removed, since these variables were found not to be important for determining cluster membership or imputing the categories introduced unnecessary bias into the sample. 

\begin{table}[htbp!]
\centering
\caption{Summary of missing values and imputation for clustering}
\label{tbl:imputation}
\begin{tabular}{c|c|l}
Variable & Percent missing & Action \\  \hline
Age & 2.2\%     &  Imputed with mean  \\
Residency & 0.47\%     &  Imputed with mode  \\
Risk tolerance & 14.16\%     &  Removed from clustering algorithm  \\
Investment objective & 6.7\%     &  Removed clients with missing information \\
Annual income & 0.13\%     &  Imputed with mean  \\
Investment knowledge level & 7.8\%     &  Removed clients with missing information  \\
Gender & 8.04\%     &  Removed clients with missing information
\end{tabular}
\end{table}
\newpage
\section*{Appendix C - Risk tolerance score distribution analysis} \label{appendixC}
In this appendix, we investigate the statistical differences between RT score distributions shown in Figure \ref{fig:riskToleranceScoresByCluster} and discussed in Section \ref{sec:withinCluster}. Table \ref{tbl:anova} shows the results of an ANOVA for RT scores where we reject the null hypothesis that the means of each cluster's RT score distribution are the same. Table \ref{tbl:tukey} shows the result of Tukey's multiple comparison test with adjusted $P$-values. The test shows that clusters 3 and 4 have significantly different means than each other and all other clusters, and clusters 1, 2, and 5 cannot reject that the means are different from each other. 
\begin{table}[hbpt]
\centering
\caption{A one-way ANOVA for comparing the means of RT scores for different clusters}
\label{tbl:anova}
\begin{tabular}{lrrrrr}
  \hline
 & Df & Sum Sq & Mean Sq & F value & Pr($>$F) \\ 
  \hline
Cluster     & 4 & 178.83 & 44.71 & 86.11 & $<0.0001$ \\ 
  Residuals   & 47556 & 24690.17 & 0.52 &  &  \\ 
   \hline
\end{tabular}
\end{table}
\begin{table}[hbpt]
\centering
\caption{Pairwise multiple comparisons using Tukey's test for the one-way ANOVA in Table \ref{tbl:anova}}
\label{tbl:tukey}
\begin{tabular}{rrrrr}
  \hline
 Clusters & Difference in means & Adjusted $P$-value \\ 
  \hline
2-1 & -0.017 &  0.345 \\ 
  3-1 & -0.074 & $<0.001$ \\ 
  4-1 & -0.247 & $<0.001$ \\ 
  5-1 & -0.008 & 0.973 \\ 
  3-2 & -0.057 & $<0.001$ \\ 
  4-2 & -0.229 & $<0.001$ \\ 
  5-2 & 0.010 & 0.900 \\ 
  4-3 & -0.172 & $<0.001$ \\ 
  5-3 & 0.067 & $<0.001$ \\ 
  5-4 & 0.239 & $<0.001$ \\ 
   \hline
\end{tabular}
\end{table}

Table \ref{tbl:kruskal} shows the results of Kruskal-Wallis test and we reject the null hypothesis in favour of at least one of the other clusters' RT score distribution stochastically dominates.  Table \ref{tbl:dunns} shows the post hoc analysis of Dunn's test, which is an analogous analysis to Tukey's test for the nonparametric setting. The results of a Dunn's test show the same result as Tukey's test, where clusters 3 and 4 pairwise stochastically dominate over the other clusters. 

\begin{table}[hbpt]
\centering
\caption{Kruskal-Wallis test for stochastic dominance of the clusters' RT score distribution.}
\label{tbl:kruskal}
\begin{tabular}{rlrrrrl}
  \hline
  & $N$ & $H$-statistic & Degrees of freedom & $P$-value \\ 
  \hline
 Cluster & 47561 & 371.93 & 4 & $<1\times10^{-79}$\\ 
   \hline
\end{tabular}
\end{table}

\begin{table}[hbpt]
\caption{Dunn's test for pairwise multiple comparisons of stochastic dominance with an adjusted $P$-value}
\label{tbl:dunns}
\centering
\begin{tabular}{rlrrrrr}
  \hline
 Cluster pair & $N_2$ & $N_2$ & Statistic & $P$-value & Adjusted $P$-value \\ 
  \hline
1-2 & 8970 & 17079 & -0.938 & 0.348 & 0.732 \\ 
  1-3 & 8970 & 12701 & -7.293 & $<0.001$ & $<0.001$ \\ 
  1-4 & 8970 & 3175 & -16.691 & $<0.001$ & $<0.001$ \\ 
  1-5 & 8970 & 5636 & 0.333 & 0.739 & 0.739 \\ 
  2-3 & 17079 & 12701 & -7.541 & $<0.001$ & $<0.001$ \\ 
  2-4 & 17079 & 3175 & -17.202 & $<0.001$ & $<0.001$ \\ 
  2-5 & 17079 & 5636 & 1.165 & 0.244 & 0.732 \\ 
  3-4 & 12701 & 3175 & -12.303 & $<0.001$ & $<0.001$ \\ 
  3-5 & 12701 & 5636 & 6.638 & $<0.001$ & $<0.001$ \\ 
  4-5 & 3175 & 5636 & 15.789 & $<0.001$ & $<0.001$ \\ 
   \hline
\end{tabular}
\end{table}

Table \ref{tbl:kldiv} shows the estimates of the symmetric KL divergences using the histogram functions in Figure \ref{fig:riskToleranceScoresByCluster} as a plug-in density estimator. These divergences represent the information lost between the two RT score distributions and measures how similar they are, where a divergence of zero means they are identically distributed. We see that clusters 1,2, and 5 distributions are very similar, where cluster 3's distribution is somewhat less similar. The most different distribution is cluster 4.  

\begin{table}[hbpt]
\centering
\caption{Symmetric KL divergence estimates for a pairwise comparison of each cluster's risk tolerance score. The left-hand column represents the distribution is being compared to the reference distribution in the first row.}
\label{tbl:kldiv}
\begin{tabular}{rl}
  \hline
 Cluster pair & Symmetric KL estimate\\ 
  \hline
1-2 & 0.0238\\ 
  1-3 & 0.0220\\ 
  1-4 & 0.0980\\ 
  1-5 & 0.0276\\ 
  2-3 & 0.0102 \\ 
  2-4 & 0.0689 \\ 
  2-5 & 0.0052\\ 
  3-4 & 0.0445\\ 
  3-5 & 0.0102\\ 
  4-5 & 0.0773 \\ 
   \hline
\end{tabular}
\end{table}
\section*{Acknowledgements}
This research was supported, in part, by funding from the Centre for Quantitative Analysis and Modelling (CQAM) at the Fields institute, and our anonymous industry partner. The authors would like to thank Adam Metzler (Wilfird Laurier University), Matt Davison (University of Western Ontario), Yuhao Zhou (Wilfrid Laurier University), Lori Weir (Four Eyes Financial), Kendall McMenamon (Four Eyes Financial), Philip Patterson (Four Eyes Financial) and the many members of our data donor team for their valuable input and insights that improved the content and writing of this document. 

\end{document}